\documentclass[preprint,12pt,2017]{elsarticle}
\usepackage{float,ltxtable}
\usepackage{multirow}
\usepackage{amssymb}
\usepackage{url}
\usepackage{tablefootnote}
\usepackage{scrextend}
\usepackage{hyperref}
\usepackage{footnote}

\journal{Icarus}

\begin{document}

\begin{frontmatter}

\title{Statistical Analysis of Astrometric Errors for the Most Productive Asteroid Surveys}

\author[jpl]{Peter Vere\v{s}}\corref{cor1}

\ead{peter.veres@cfa.harvard.edu}
\author[jpl]{Davide Farnocchia}
\author[jpl]{Steven R. Chesley}
\author[jpl]{Alan B. Chamberlin}

\address[jpl]{Jet Propulsion Laboratory, California Institute of Technology, 4800 Oak Grove Drive, Pasadena, CA, 91109}
\cortext[cor1]{Present address: {\em Minor Planet Center, Harvard-Smithsonian Center for Astrophysics, 60 Garden Street, Cambridge, MA 02138}}

\begin{abstract}
We performed a statistical analysis of the astrometric errors for the major asteroid surveys. We analyzed the astrometric residuals as a function of observation epoch, observed brightness and rate of motion, finding that astrometric errors are larger for faint observations and some stations improved their astrometric quality over time. Based on this statistical analysis we develop a new weighting scheme to be used when performing asteroid orbit determination. The proposed weights result in ephemeris predictions that can be conservative by a factor as large as 1.5. However, the new scheme is robust with respect to outliers and handles the larger errors for faint detections. \end{abstract}


\end{frontmatter}

\section{Introduction}

The first discovery of a minor planet (Ceres) in 1801 by Giuseppe Piazzi was possible due to its recovery based on a novel orbit determination method derived by \citet{Gauss1809}. Since then, observational techniques for detecting asteroids have shifted from visual to photographic and photography was replaced by the charge-coupled device (CCD). Due to their high sensitivity, fast readout and use of computer algorithms, CCDs have provided by far the largest amount of positional astrometry for known asteroids. With the new observational methods, measurement uncertainties decreased, thus improving the quality of asteroid orbits.

The first generation of asteroid surveys had the goal of discovering 90\% of Near Earth Objects (NEOs) larger than 1\,km as mandated by the US government to NASA in 1998. The emerging large-format CCDs, dedicated observing time and NASA funding led to a dramatic increase in the discovery rate of asteroids and comets. The first generation of dedicated surveys in 1995-2010 was represented by Spacewatch \citep{McMillan06} on Kitt Peak in Arizona; Lincoln Near-Earth Asteroid Research \citep[LINEAR,][]{Stokes00} in Socorro, New Mexico; Lowell Observatory NEO Survey \citep[LONEOS,][]{Koehn99} in Flagstaff, Arizona; and Near-Earth Object Tracking \citep[NEAT,][]{Pravdo99} on Haleakala, Hawaii and at Palomar, California.  In 2005, the Catalina Sky Survey \citep{Larson03}, which included sites at Mt. Bigelow (Catalina) and Mt. Lemmon near Tucson, Arizona, and Siding Spring in Australia, became a major contributor. 

Pan-STARRS1 \citep{Hodapp04} is an example of next generation all-sky telescopes focusing on multiple fields of astronomy. In 2011 Pan-STARRS1 started its three-year operations on Haleakala, Maui and in 2014 it turned into a dedicated NEO survey, thus becoming the major contributor of  NEO discoveries. Also, the Space Surveillance Telescope on Atom Peak in New Mexico has been submitting large quantities of incidental asteroid astrometry for the last few years. Additionally, the WISE infrared space telescope and its NEO component NEOWISE \citep{Wright10} contributed significantly to the discovery of asteroids and comets. 

According to NEO population models \citep{Harris15}, we have reached an estimated discovery completeness of $95\%$ for NEOs larger than 1\,km NEOs and of $30\%$ for NEOs larger 140\,m. In the near future, LSST \citep{2016IAUS..318..282J} and possibly NEOCAM \citep{2016AJ....151..172G} will increase the number of known asteroids by an order of magnitude.

The reliability of asteroid orbits significantly depends on the quality and the statistical treatment of astrometric observations. One way to improve the astrometric precision is to use large telescopes, with a long focal length, good pixel resolution, and located at observing sites with excellent seeing. However, access to large telescopes is rather limited and expensive, the field of view is  usually very small and ground-based astrometry can be limited by seeing. Nevertheless, orbits can be improved by extending the data arcs, possibly with archival observations, by removing known biases introduced by the reference catalogs, and by adopting weighting schemes that account for the diverse quality of the astrometric dataset.

\citet{Carpino03} first studied the problem of existing systematic errors in asteroid astrometry due to star catalog biases. \citet{Chesley10} computed systematic errors for multiple star catalogs in comparison to 2MASS \citep{Skrut06}, and provided correction tables that lowered the systematic errors and improved the accuracy of the computed orbits. \citet{Farnocchia15} refined the star catalog debiasing tables by including the effect of star proper motions and extending the scheme to a larger set of catalogs.

\citet{Chesley10} and \citet{Farnocchia15} also proposed weighting schemes based on the RMS for individual stations, eventually grouping by star catalog those stations with lower number of observations. However, in reality the astrometric quality was not always the same throughout the operation of individual surveys. Often telescopes were upgraded, with increased detection sensitivity, improved astrometric reduction pipeline or a different reference frame star catalog. Even human factors could affect the quality of the submitted data.
 
The availability of a reliable weighting scheme is especially important since the current observation format of the Minor Planet Center does not include astrometric uncertainties, which then need to be assumed when computing asteroid orbits.
Therefore, the scope of this work is to statistically analyze the residuals of debiased astrometric positions from major asteroid CCD surveys as a function of reported magnitude, epoch and type of observation. Ultimately, based on the derived statistics we present a new weighting scheme. The new scheme is now being used by the JPL Solar System Dynamics orbit determination pipeline, including the newly released Scout system \citep{Farnocchia15a, Farnocchia16}.

%


\section{Statistical analysis of asteroid CCD astrometry}

As of February 9, 2016 the asteroid CCD astrometry dataset contains more than 130 million astrometric measurements of about 700,000 asteroids (Table~\ref{surveys}). The 13 most productive asteroid CCD surveys produced more than 91\% of this dataset and so we focused our analysis on fully characterizing them. The positional astrometry of this dataset is debiased with respect to known star catalog biases by \citet{Farnocchia15} and the residuals computed by the JPL orbit determination pipeline. To make sure the observation residuals reflected the actual astrometric errors, we restricted our analysis to the astrometric dataset corresponding to multi-apparition orbits. 

\begin{table*}[ht!]
\small
\begin{center}
\caption{The most productive asteroid surveys are listed with the RMS of their astrometric residuals for multi-apparition asteroids. Statistics for all detections and the fraction of observed known asteroids is current as of February 9, 2016.}
\begin{tabular}{c|c|c|cc|cc|c}
\hline
\multirow{2}{*}{N} &Station &MPC & RA &  DEC&  \multirow{2}{*}{Detections} & \multirow{2}{*}{Asteroids} & Fraction of \\
 & Name&Code &RMS &RMS & &&Asteroids \\
\hline
1 &LINEAR&704&0.67$''$&0.66$''$&32,777,288&370,033&53\%\\
2 &Mt. Lemmon&G96&0.31$''$&0.28$''$&18,640,225&619,386&88\%\\
3 &Pan-STARRS1&F51&0.12$''$&0.12$''$&18,400,219&616,209&88\%\\
4 &Catalina&703&0.69$''$&0.67$''$&17,802,653&436,810&62\%\\
5&Spacewatch&691&0.37$''$&0.34$''$&11,719,895&566,880&81\%\\
6&SST&G45&0.36$''$&0.36$''$&9,915,512&300,495&43\%\\
7&LONEOS&699&0.65$''$&0.59$''$&5,367,447&261,585&37\%\\
8&NEAT&644&0.30$''$&0.36$''$&3,926,121&302,846&43\%\\
9&Purple Mountain&D29&0.50$''$&0.47$''$&3,214,197&274,301&39\%\\
10&WISE&C51&0.55$''$&0.59$''$&2,222,396&149,884&21\%\\
11&Siding Spring&E12&0.49$''$&0.52$''$&2,228,965&168,462&24\%\\
12&Haleakala-AMOS&608&0.72$''$&0.85$''$&1,286,280&144,827&21\%\\
13&La Sagra&J75&0.42$''$&0.39$''$&1,159,632&153,233&22\%\\
\hline
\end{tabular}
\label{surveys}
\end{center}
\end{table*}

The first measure to assess the data quality is to compute the root-mean-square (RMS) of astrometric residuals in RA and DEC (Table~\ref{surveys}). Among surveys Pan-STARRS1 achieves the best accuracy with an RMS of 0.12$''$. Yet, the RMS might not capture outliers, trends like time dependence, signal-to-noise ratio (SNR), brightness and rate of motion, which we analyze in the following subsections.

\subsection{Astrometric residuals as a function of observation epoch}\label{sec:res_time}

Over time surveys experience changes in the star catalog used, upgraded CCD cameras and telescopes, and improved measuring algorithms. All of these changes can affect the astrometric quality. The top panels of Figures~\ref{fig.rms1}--\ref{fig.rms6} and the right panel of Figure~\ref{fig.rms7} show the dependency on epoch of observation for the considered surveys. In some cases the observation epoch matters (LINEAR, Catalina, Spacewatch, NEAT, AMOS) in other cases it does not seem to be a factor. For instance, the Spacewatch survey started using a new mosaic camera in late 2002 and saw an improvement of the astrometry in the following years. On the contrary, Pan-STARRS1 has been using the same catalog, same instrument and same reduction and automated pipeline, therefore no change over time was visible. 

\subsection{Astrometric residuals as a function of target brightness}

\begin{figure}[H]
  \centering
             \includegraphics[width=0.48\textwidth]{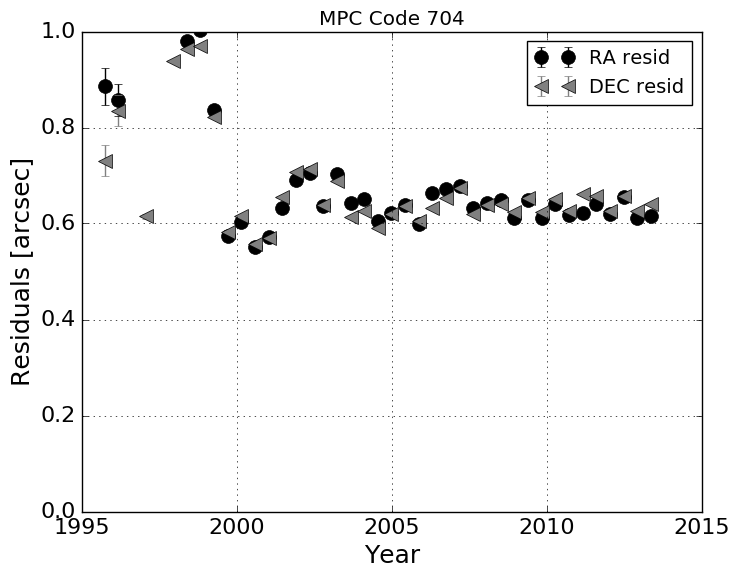}
    \includegraphics[width=0.48\textwidth]{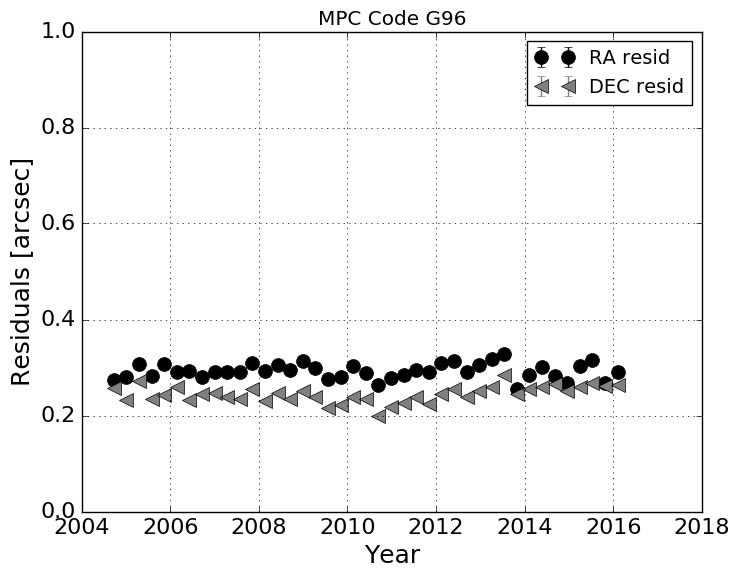}
       \includegraphics[width=0.48\textwidth]{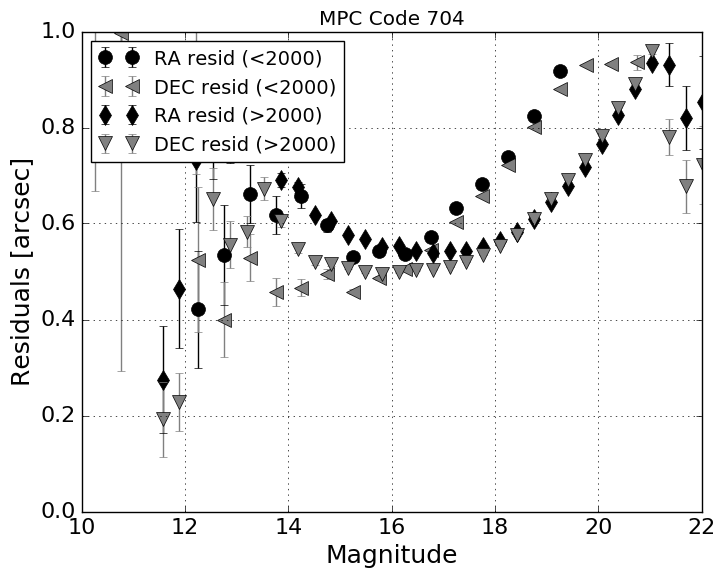}
    \includegraphics[width=0.48\textwidth]{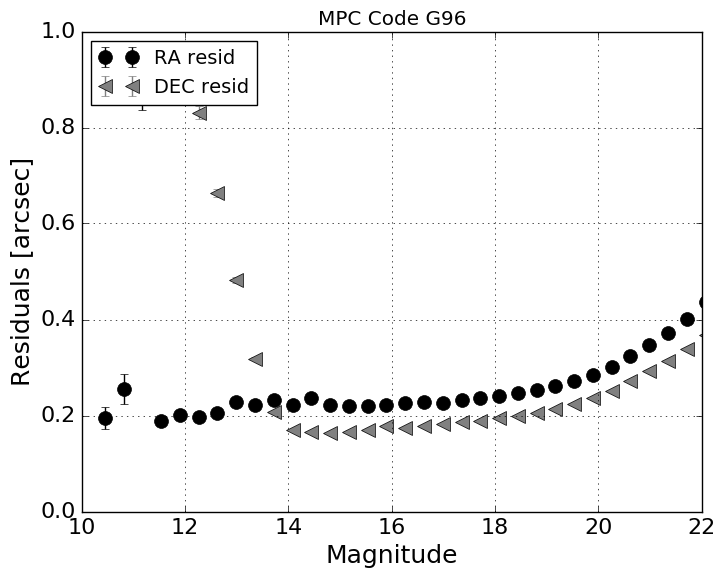}
           \includegraphics[width=0.48\textwidth]{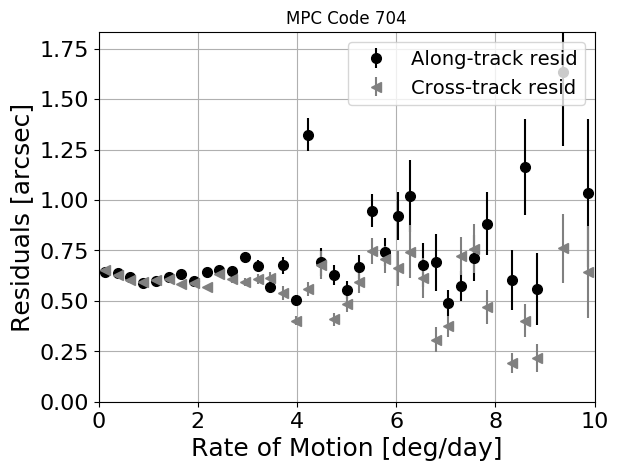}
    \includegraphics[width=0.48\textwidth]{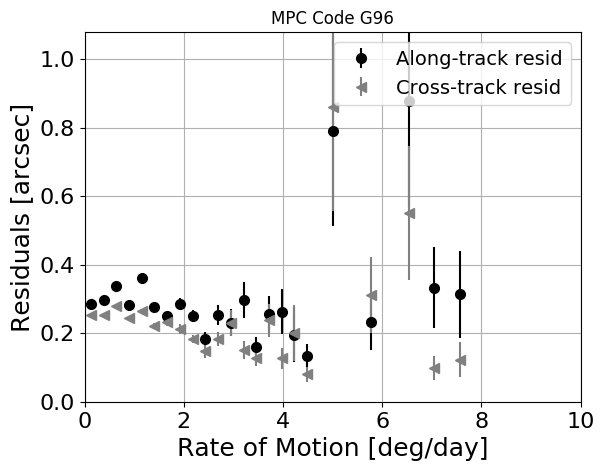}
     \caption{RMS of astrometric residuals of multi-apparition asteroids in right ascension and declination (top and middle panels) and along-track and cross-track astrometric residuals (bottom panels) as a function of epoch (top panels), magnitude (middle panels) and rate of motion (bottom panels) for LINEAR (left panels) and the Mt. Lemmon Survey (right panels).}
    \label{fig.rms1}
\end{figure}

\begin{figure}[H]
  \centering
            \includegraphics[width=0.48\textwidth]{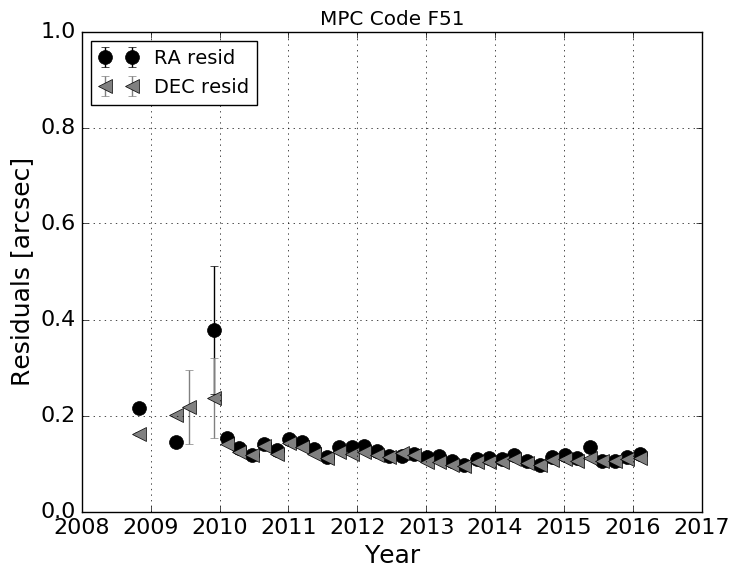}
    \includegraphics[width=0.48\textwidth]{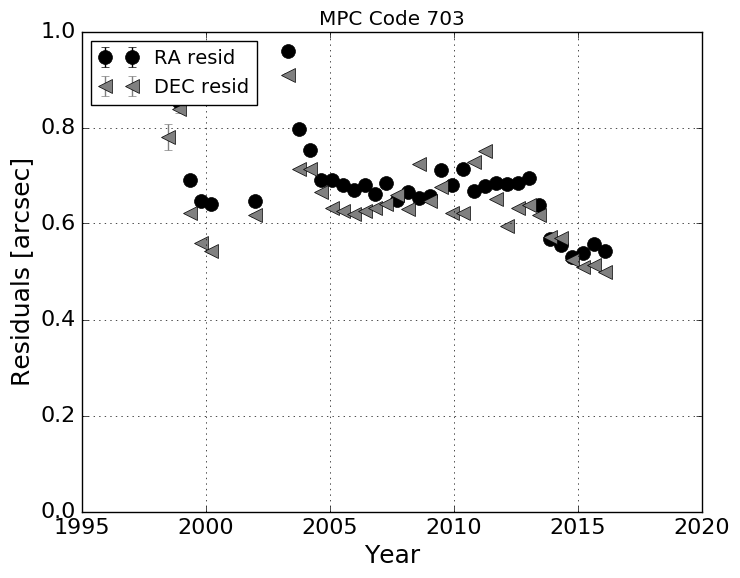}
       \includegraphics[width=0.48\textwidth]{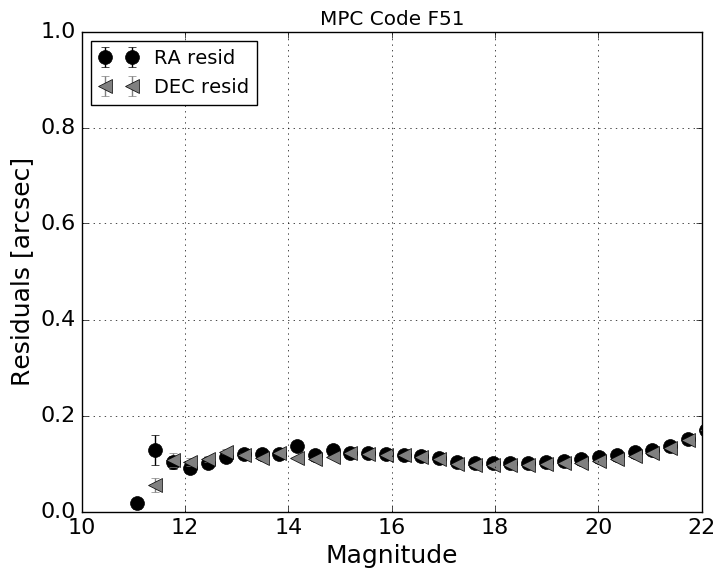}
    \includegraphics[width=0.48\textwidth]{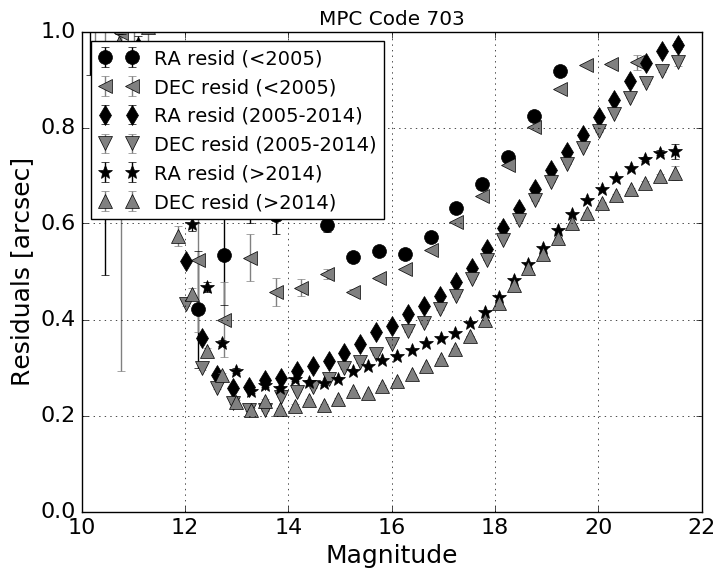}
           \includegraphics[width=0.48\textwidth]{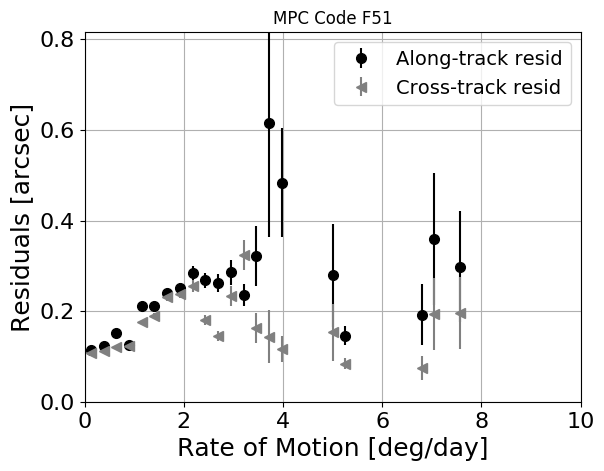}
    \includegraphics[width=0.48\textwidth]{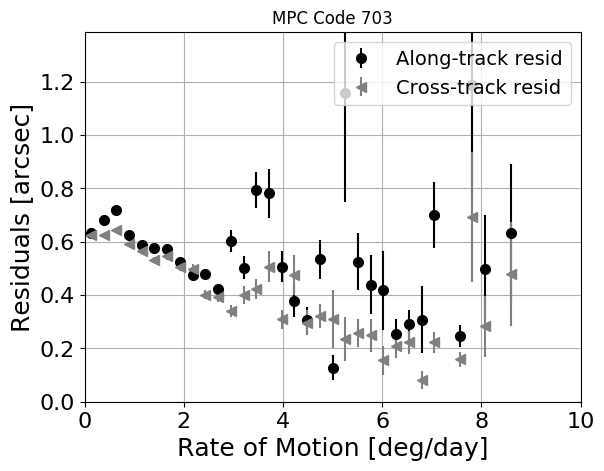}
     \caption{RMS of astrometric residuals of multi-apparition asteroids in right ascension and declination (top and middle panels) and along-track and cross-track astrometric residuals (bottom panels) as a function of epoch (top panels), magnitude (middle panels) and rate of motion (bottom panels) for Pan-STARRS (left panels) and the Catalina Sky Survey (right panels).}
    \label{fig.rms2}
\end{figure}

\begin{figure}[H]
  \centering
             \includegraphics[width=0.48\textwidth]{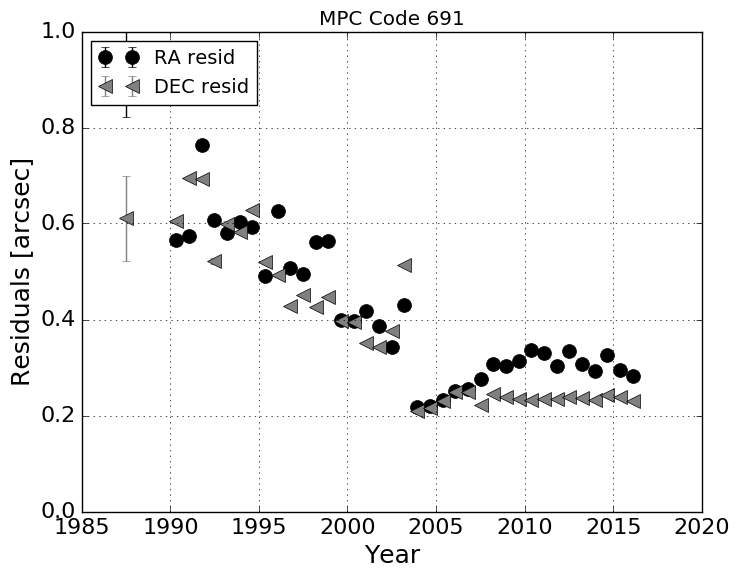}
    \includegraphics[width=0.48\textwidth]{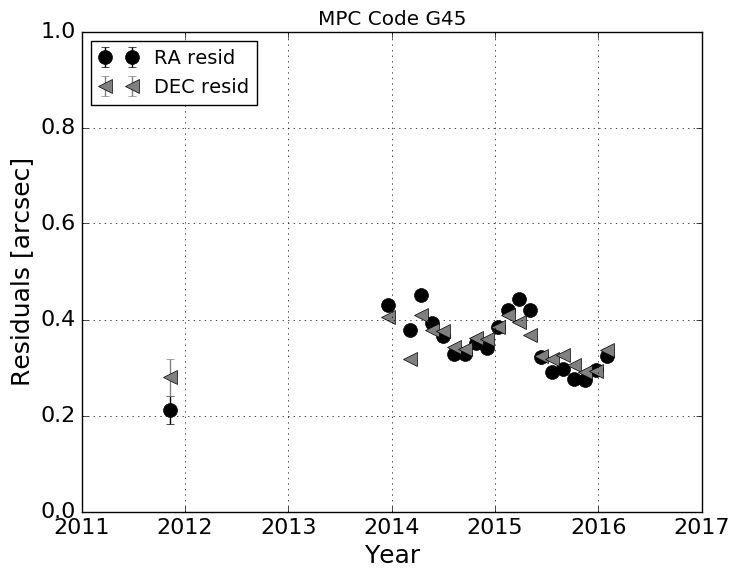}
       \includegraphics[width=0.48\textwidth]{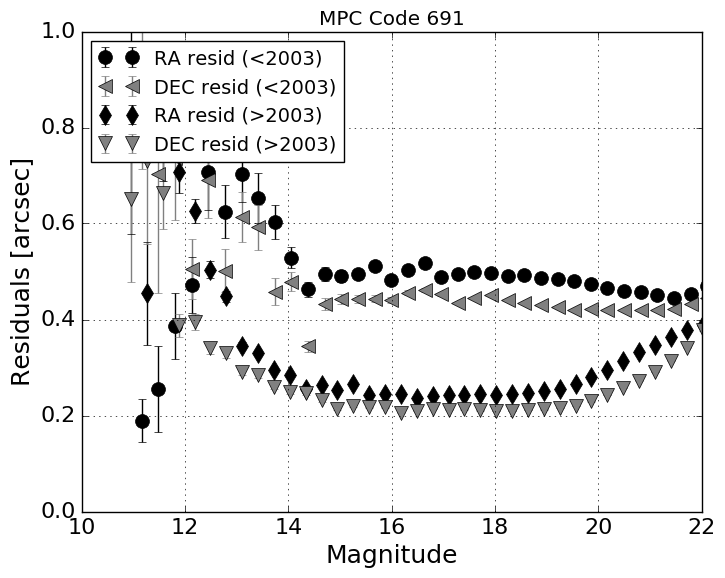}
    \includegraphics[width=0.48\textwidth]{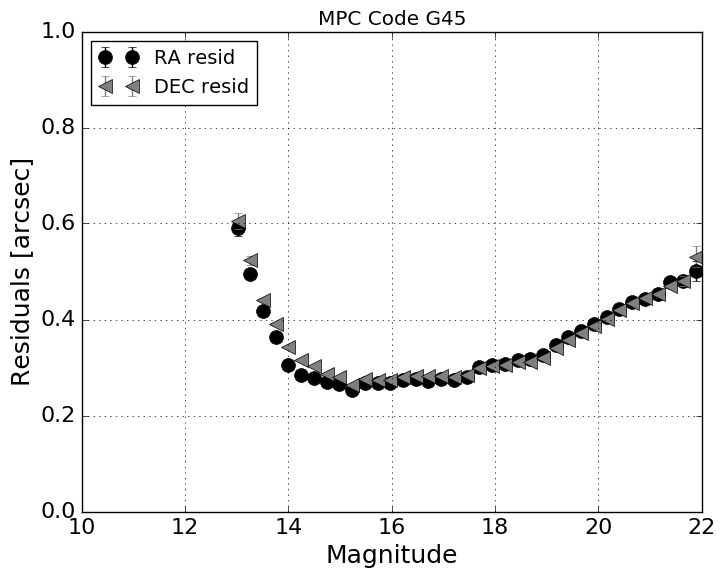}
           \includegraphics[width=0.48\textwidth]{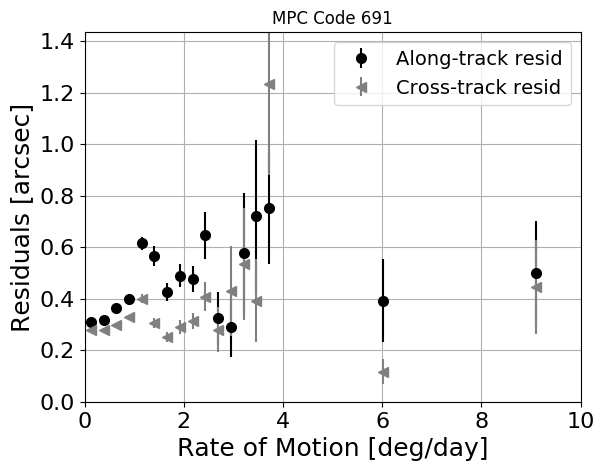}
    \includegraphics[width=0.48\textwidth]{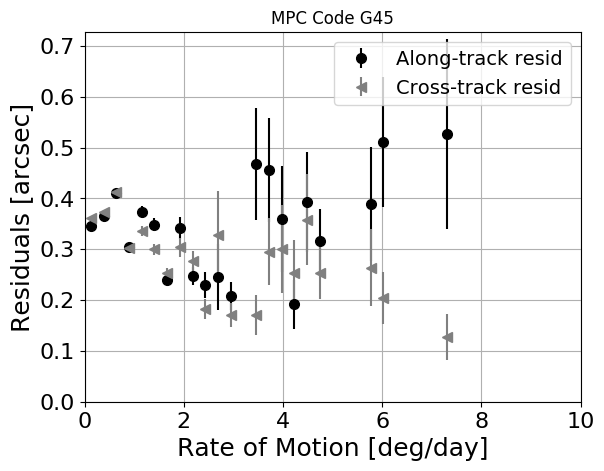}
     \caption{RMS of astrometric residuals of multi-apparition asteroids in right ascension and declination (top and middle panels) and along-track and cross-track astrometric residuals (bottom panels) as a function of epoch (top panels), magnitude (middle panels) and rate of motion (bottom panels)) for Spacewatch (left panels) and the Space Surveillance Telescope (right panels).}
    \label{fig.rms3}
\end{figure}

\begin{figure}[H]
  \centering
            \includegraphics[width=0.48\textwidth]{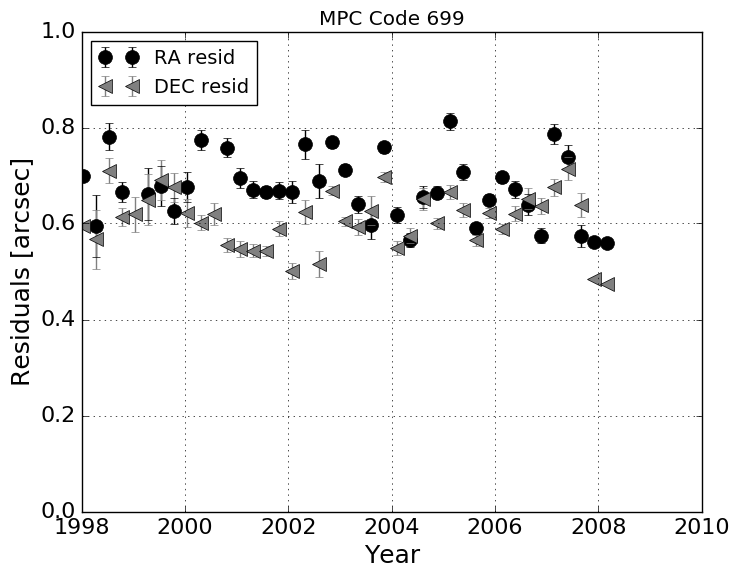}
    \includegraphics[width=0.48\textwidth]{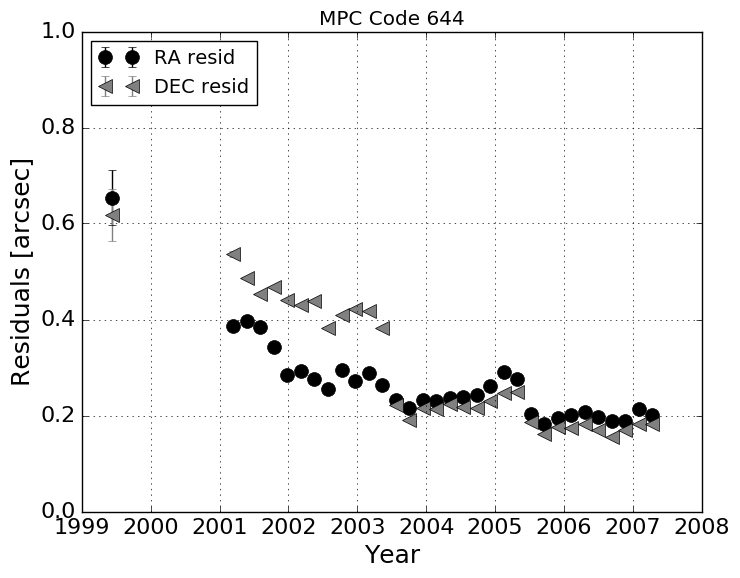}
       \includegraphics[width=0.48\textwidth]{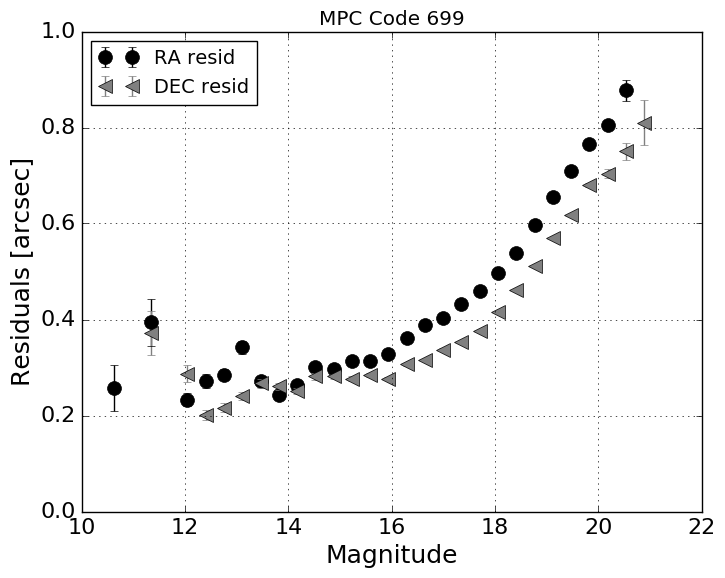}
    \includegraphics[width=0.48\textwidth]{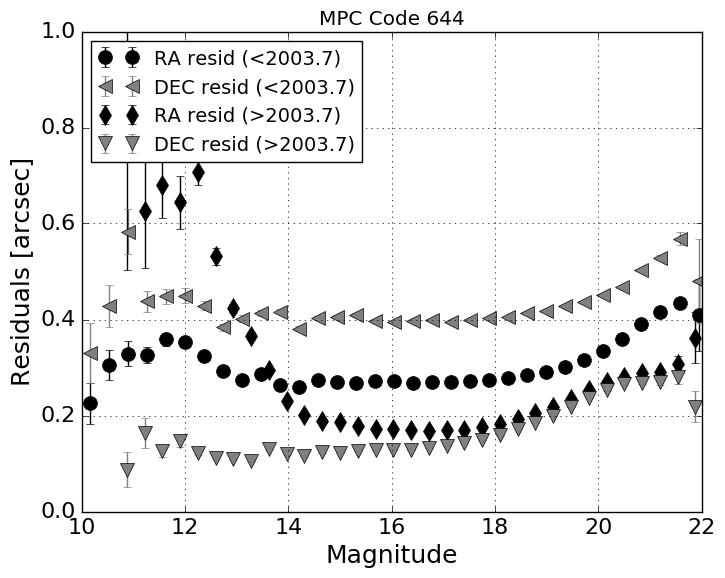}
           \includegraphics[width=0.48\textwidth]{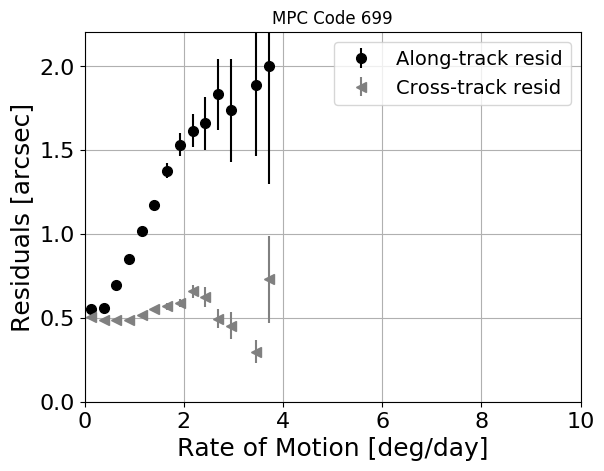}
    \includegraphics[width=0.48\textwidth]{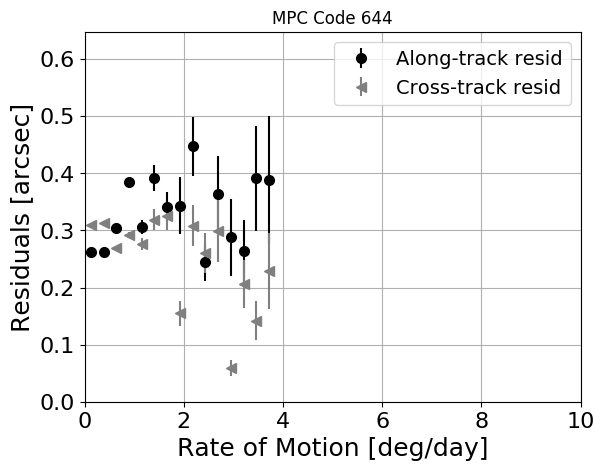}
     \caption{RMS of astrometric residuals of multi-apparition asteroids in right ascension and declination (top and middle panels) and along-track and cross-track astrometric residuals (bottom panels) as a function of epoch (top panels), magnitude (middle panels) and rate of motion (bottom panels) for LONEOS (left panels) and NEAT (right panels).}
    \label{fig.rms4}
\end{figure}

\begin{figure}[H]
  \centering
            \includegraphics[width=0.48\textwidth]{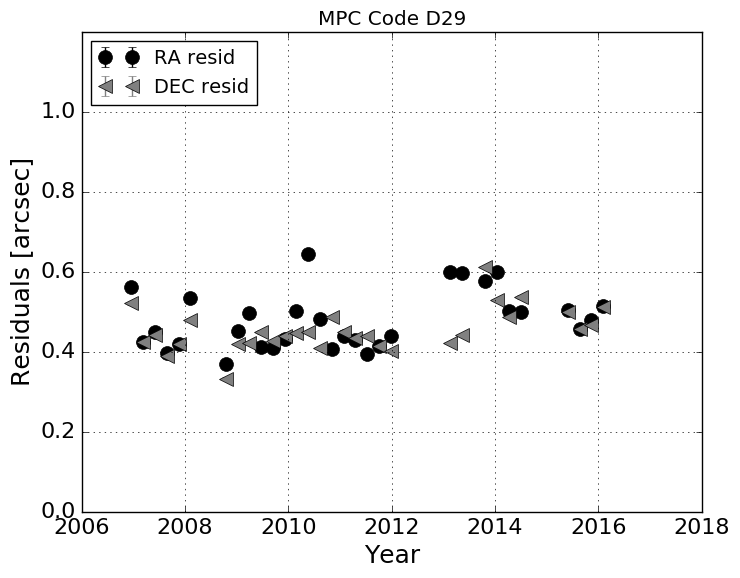}
    \includegraphics[width=0.48\textwidth]{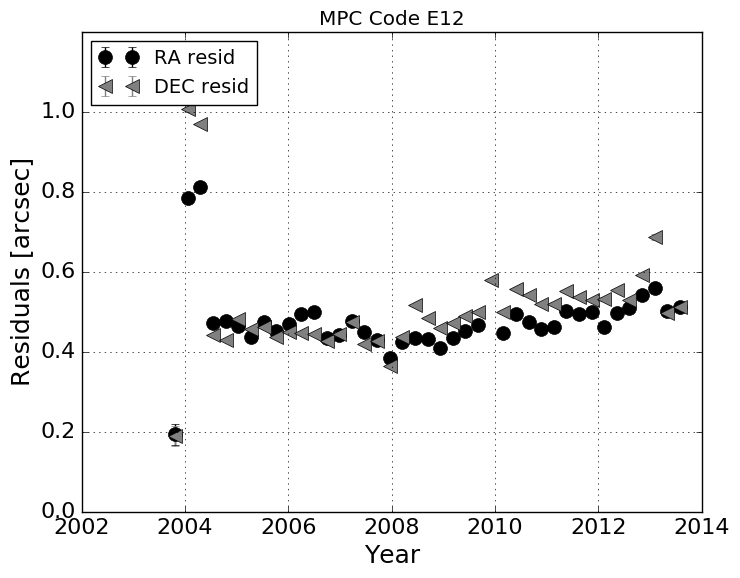}
       \includegraphics[width=0.48\textwidth]{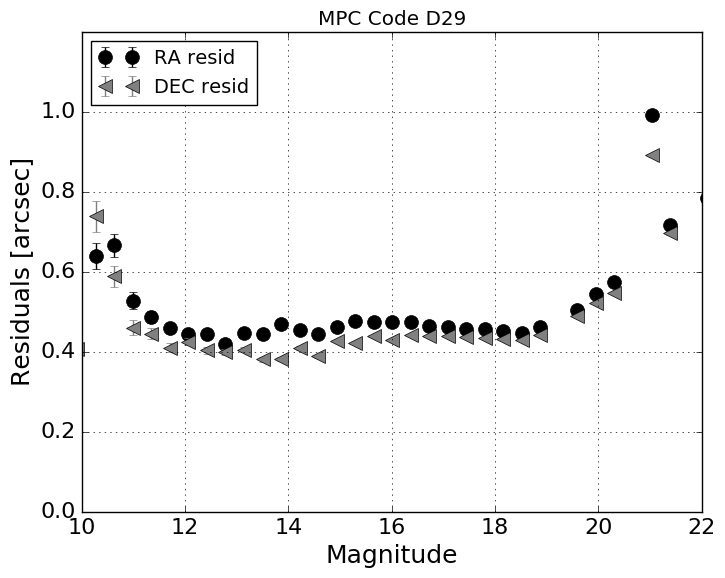}
    \includegraphics[width=0.48\textwidth]{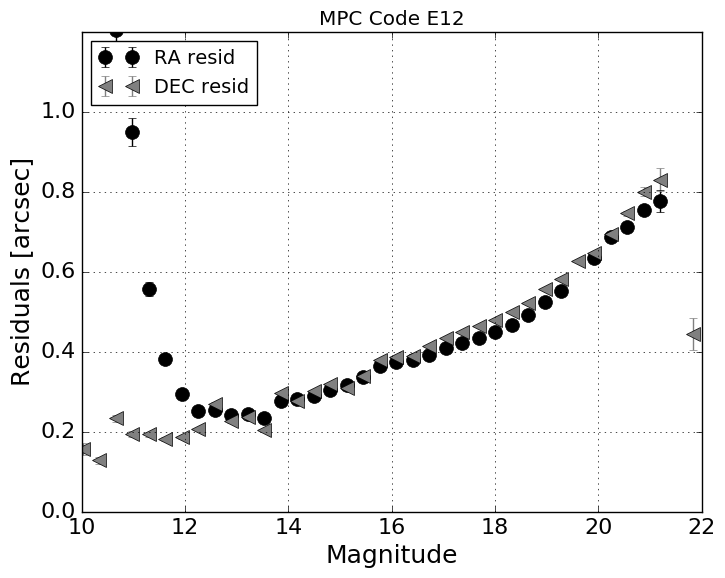}
           \includegraphics[width=0.48\textwidth]{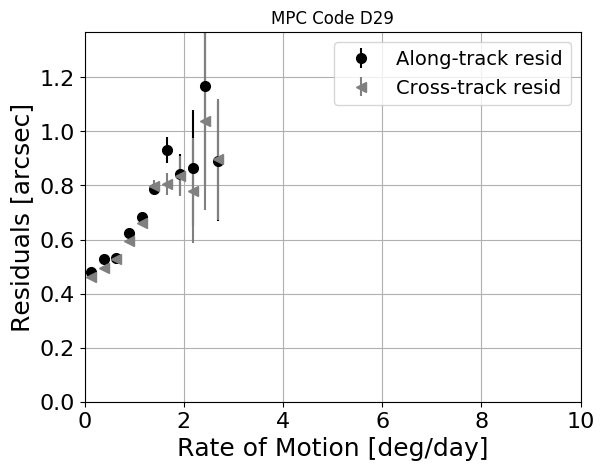}
    \includegraphics[width=0.48\textwidth]{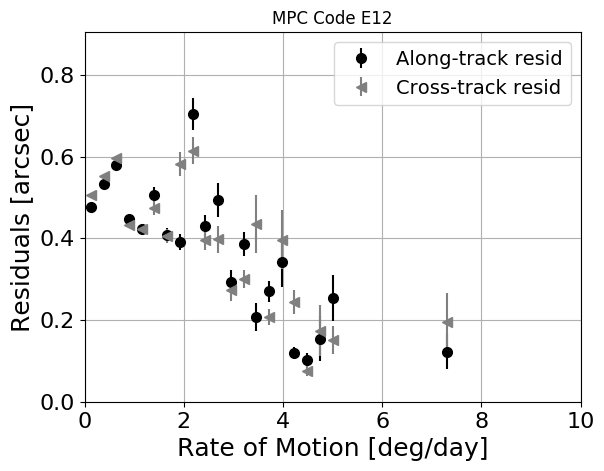}
     \caption{RMS of astrometric residuals of multi-apparition asteroids in right ascension and declination (top and middle panels) and along-track and cross-track astrometric residuals (bottom panels) as a function of epoch (top panels), magnitude (middle panels) and rate of motion (bottom panels) for Purple Mountain (left panels) and the Siding Spring Survey (right panels).}
    \label{fig.rms5}
\end{figure}

\begin{figure}[H]
  \centering
            \includegraphics[width=0.48\textwidth]{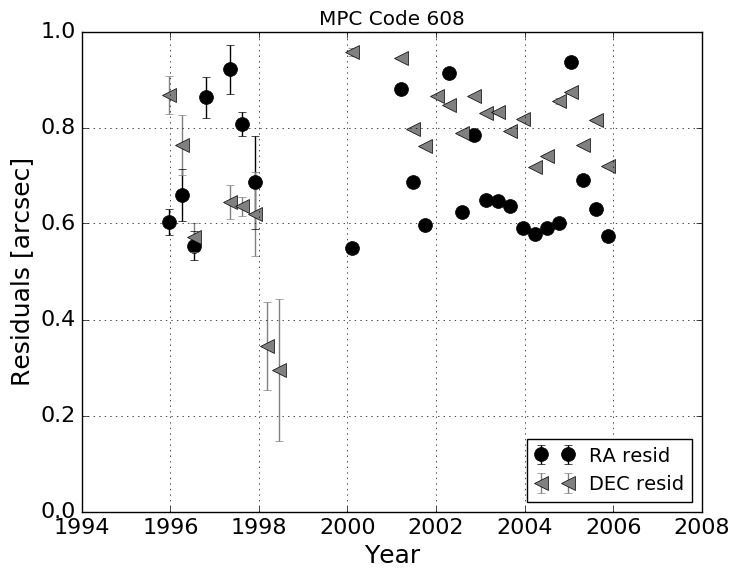}
    \includegraphics[width=0.48\textwidth]{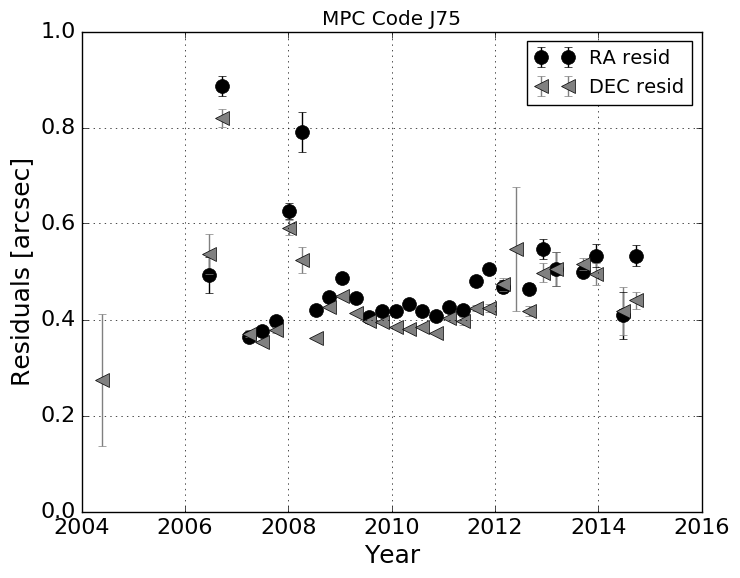}
       \includegraphics[width=0.48\textwidth]{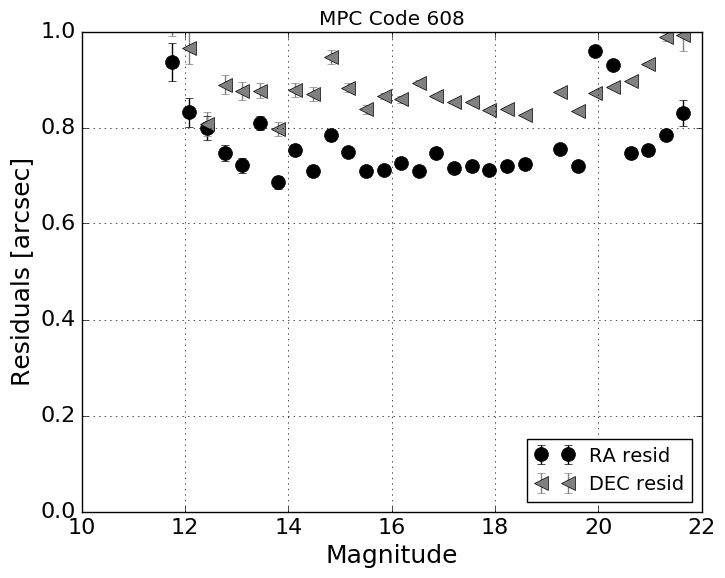}
    \includegraphics[width=0.48\textwidth]{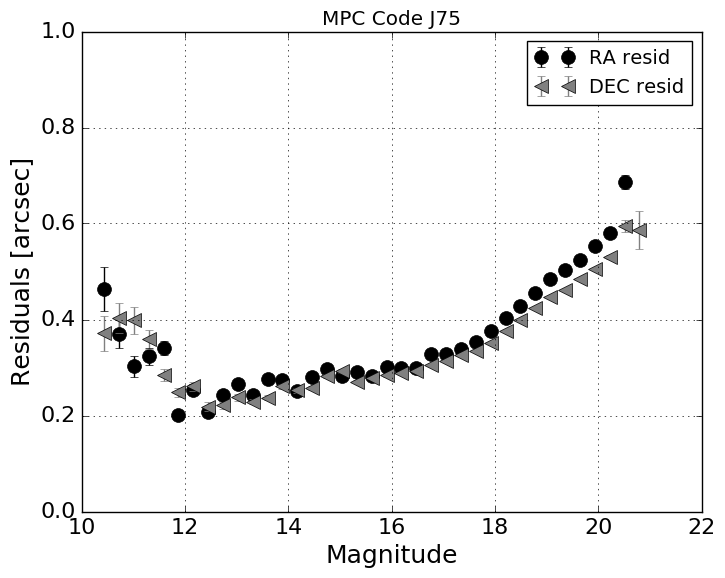}
            \includegraphics[width=0.48\textwidth]{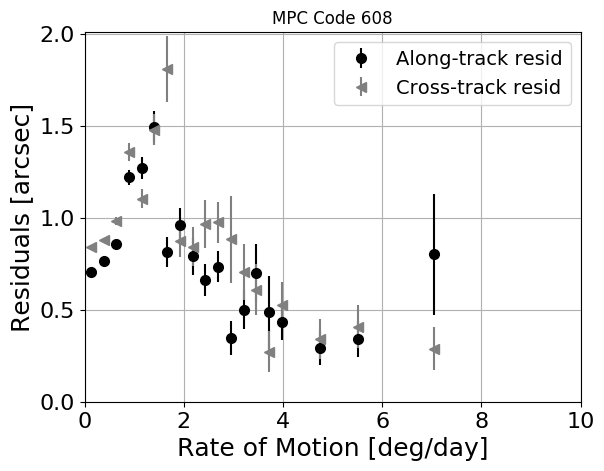}
    \includegraphics[width=0.48\textwidth]{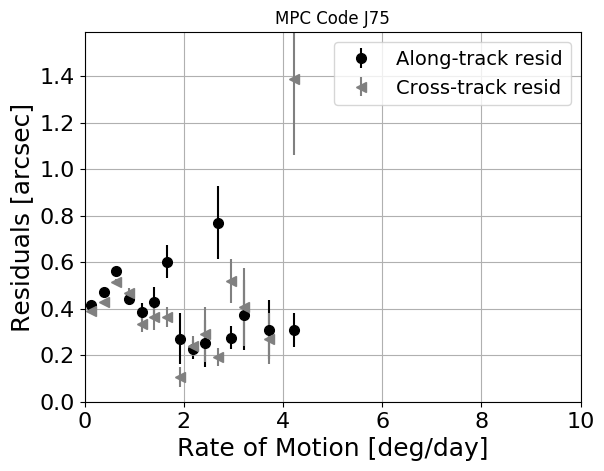}
     \caption{RMS of astrometric residuals of multi-apparition asteroids in right ascension and declination (top and middle panels) and along-track and cross-track astrometric residuals (bottom panels) as a function of epoch (top panels), magnitude (middle panels) and rate of motion (bottom panels) for Haleakala-AMOS (left panels) and La Sagra (right panels).}
    \label{fig.rms6}
\end{figure}

\begin{figure}[H]
  \centering
      \includegraphics[width=0.48\textwidth]{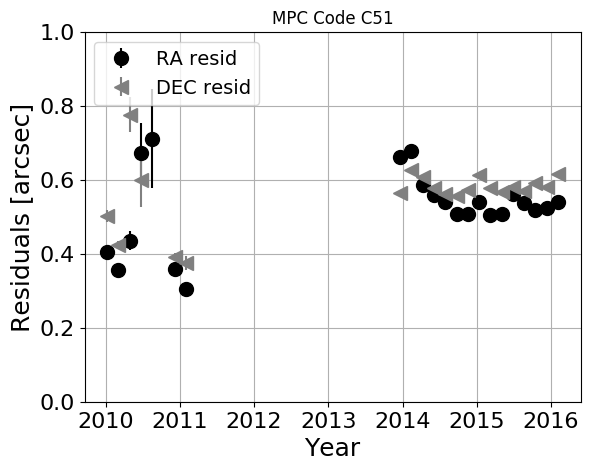}
         \includegraphics[width=0.48\textwidth]{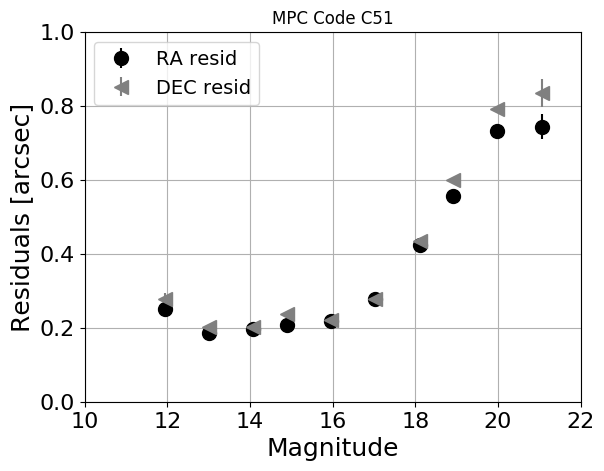}
     \caption{RMS of astrometric residuals of multi-apparition asteroids in right ascension and declination as a function of epoch (left panel) and magnitude (right panel) for WISE.}
    \label{fig.rms7}
\end{figure}

The middle panels of Figures~\ref{fig.rms1}--\ref{fig.rms6} and the left panel of Figure~\ref{fig.rms7} show the astrometric RMS as a function of the magnitude for different surveys. Astrometric errors can significantly increase with respect to the mean RMS (Table~\ref{surveys}). The astrometric uncertainty depends on SNR, which is not available because of the restrictions of the MPC astrometric format. As a proxy for SNR we can use the target's visible brightness. Faint objects have lower SNR and so uncertainties increase as we can see in the figures. Objects that are too bright are also saturated on CCDs and could provide bad photometry and astrometry.

To capture all astrometric measurements within a projected uncertainty, the proposed weights should be larger than the mean RMS and capture the data quality dependence on brightness.

\subsection{Timing and dependency on the rate of motion}

The motion of asteroids can cause detections to be trailed along the direction of motion, often increasing the measurement error along the direction of motion. When objects move fast in the plane-of-sky, there is also the possibility that timing errors introduce significant positional errors. The bottom panels of Figures~\ref{fig.rms1}--\ref{fig.rms6} show the along-track (AT) and cross-track (CT) residual RMS as a function of magnitude and time for the ground-based observatories in Table~\ref{surveys}. We computed the AT and CT-residuals by projecting the RA and DEC components to the sky-plane motion. CT should reflect the astrometric position error, while AT will also contain errors coming from clock errors or measurement errors for fast trailed asteroids. 

We expected to see larger AT errors for fast moving targets. Trailed asteroid detections tend to have larger uncertainties along-track, because the measurement errors are usually correlated with the trail length, due to point-spread-function fitting of trails instead of a proper trail representation \citep{Veres12,Fraser16}. In the case of Pan-STARRS1, LONEOS and Purple Mountain this trend is confirmed. However, in some cases the along-track residuals decrease as a function of the rate of motion (Siding Spring, Mt. Lemmon, Catalina, NEAT), remain flat (LINEAR, Spacewatch, NEAT) or have an upward and downward direction (SST). An upward trend suggests measurement errors caused by trailed detections or timing errors, while a flat trend shows that there is no timing or measurement problem for trails. Finally, a downward trend suggests that measurers might have taken special care in measuring fast moving targets.

Timing errors can be inferred by dividing the along-track residual by the rate of motion. However, it is important to point out it is not always possible to separate timing errors from position errors, especially for slow movers. 
If the timing error is large for slow moving asteroids but not for the fast ones, this could likely indicate star catalog bias.

\begin{figure}[ht!]
  \centering
        \includegraphics[width=0.8\textwidth]{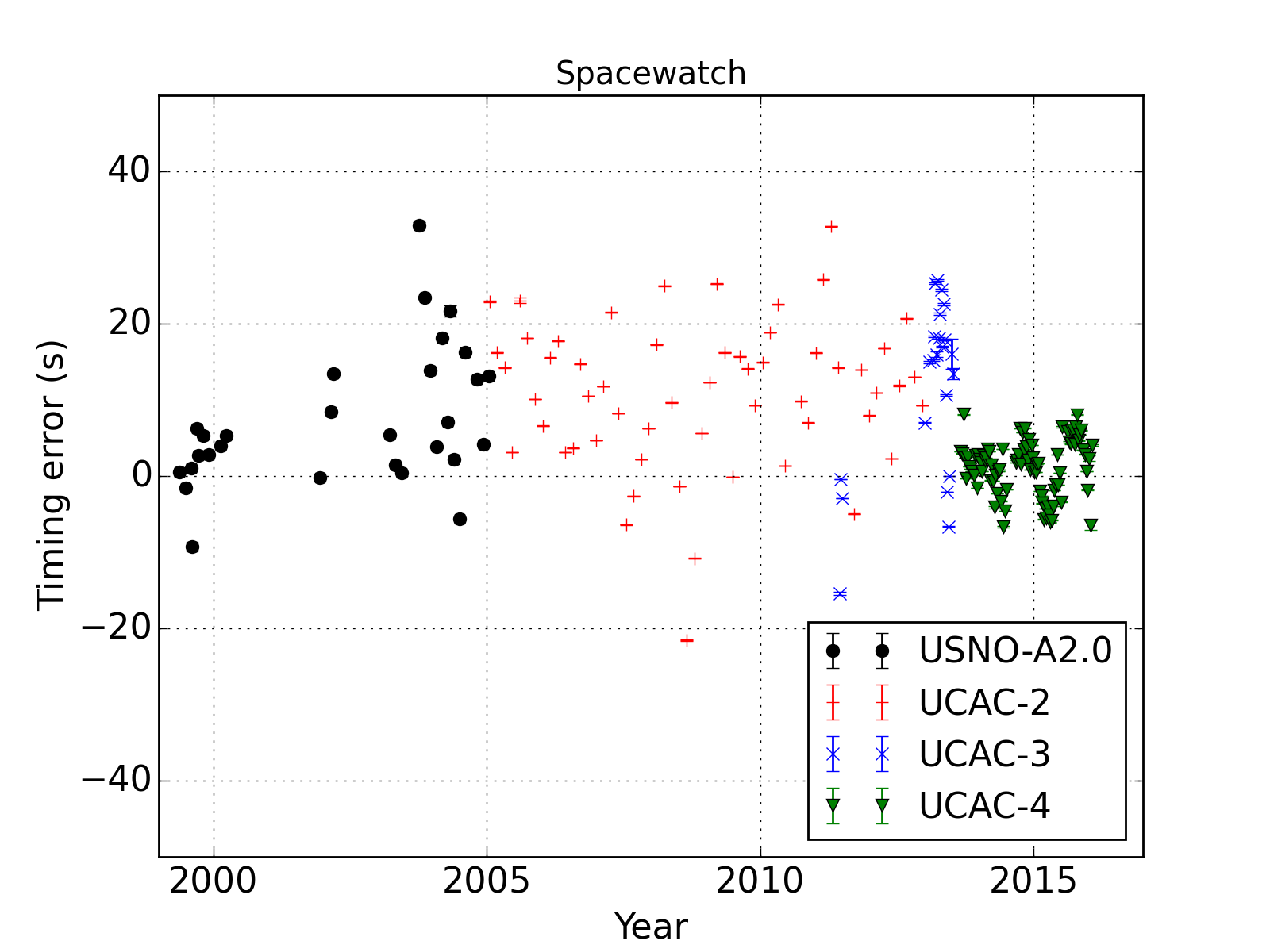}
    \includegraphics[width=0.8\textwidth]{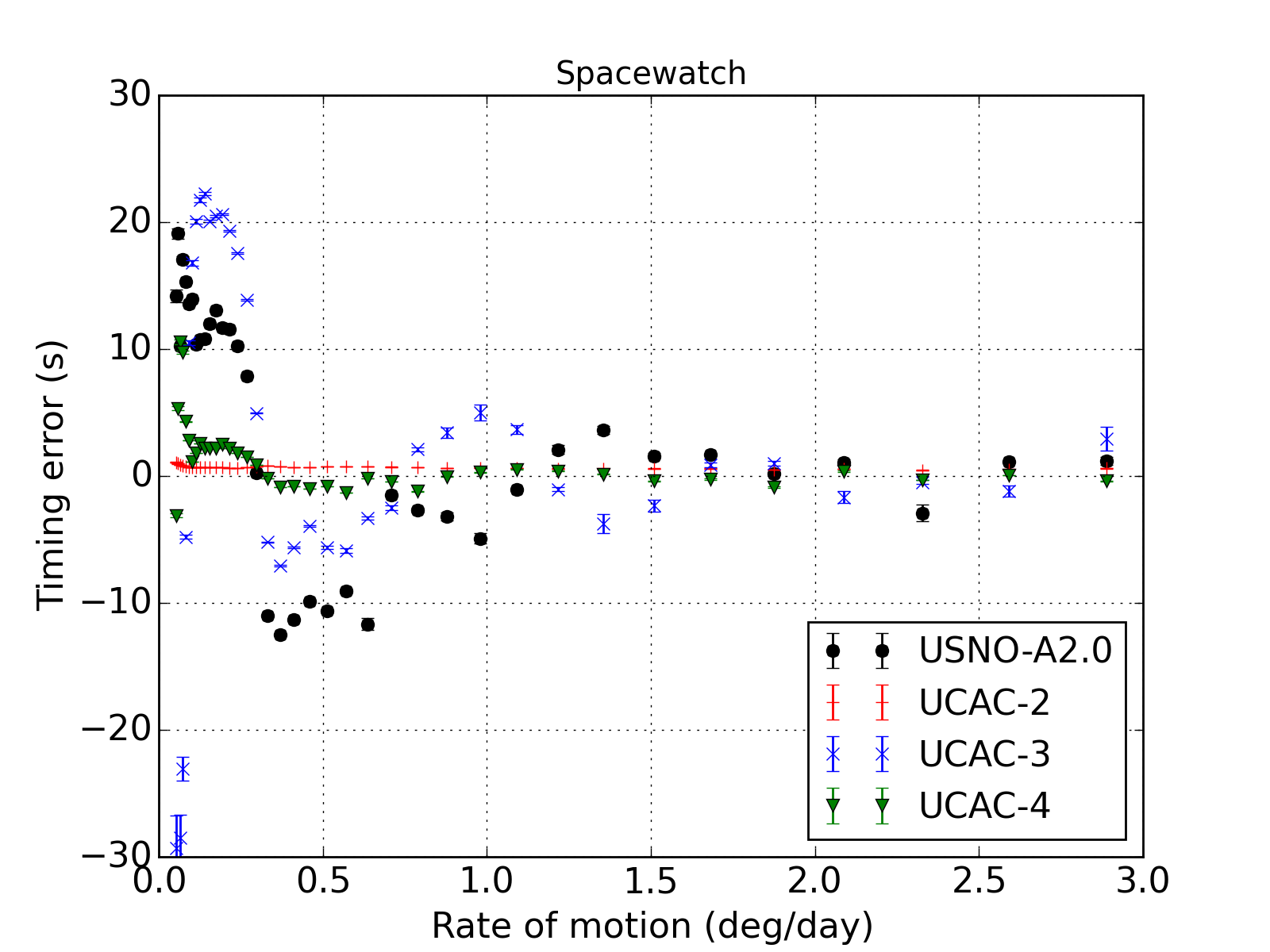}
 \caption{Timing error of Spacewatch astrometry as a function of epoch (top panel) and rate of motion (bottom panel) for star reference catalogs.}
\label{vel_2}
\end{figure}

For instance, Figure~\ref{vel_2} is an example of timing error as a function of rate of motion and epoch for Spacewatch, which used five different reference catalogs over time. Until 2014, the timing error seemed to be large, at a level of tens of seconds. However, the lower figure shows that only slow movers represented by main belt asteroids (MBAs) moving at $0.2$\,deg/day on average, are responsible for this effect and only for USNO-A2.0 and UCAC-3 catalogs. Moreover, it is clear how the time bias is catalog dependent, thus pointing to unresolved systematics in the catalogs. At this rate of motion, a catalog bias of 0.1$''$ corresponds to a timing error of 12 seconds. The lowest timing errors were visible for the UCAC-4 catalog, but even there,  a seasonal effect is clearly  visible, possibly coming from unresolved catalog biases at peculiar values of right ascension. The variation around zero is presumably correlated with the location of the ecliptic with respect to the equator when fields were observed. In general it is hard to identify timing issues for slow movers, as systematic errors in position are dominant.


\clearpage

\section{The weighting scheme for CCD observations}

The current weighting scheme used at JPL is described by \citet{Chesley10} and \citet{Farnocchia15}. The data weights are assigned based on station-specific statistics when enough observations are available, otherwise observations are binned by catalog. The weights are defined as $w=1/\sigma^2$, where $\sigma$ is the observation uncertainty, which is scaled from the RMS of the residuals. To mitigate possible effects of unresolved systematics, uncertainties are inflated by a factor of $\sqrt{N}$, where N represents the number of positions in a nightly batch from the same station. 

Based on a few years of experience, we found some issues and room for improvement to the current scheme.  In particular, on multiple occasions that scheme forced us to set manual weights for specific objects, especially when dealing with NEOs. As shown in Figures~\ref{fig.rms1}--\ref{fig.rms7}, faint detections end up being overweighted when the data weight is based on the RMS. Ideally, we should use the SNR to scale the RMS, but the SNR is not yet available. Therefore, in the scheme proposed in this paper, we conservatively set the data weights according to the upper bound of the RMS as a function of brightness. 

For some stations the astrometric quality is not uniform over time, as discussed in Section~\ref{sec:res_time}. In those cases we split in to different time intervals and analyzed the brightness dependence for each time interval. 

For observers without reliable station-specific statistics, the current weighting bins observations and assigns weights by the star catalog used for the astrometric reduction.
However, there can be a significantly inhomogeneous quality between different observers using the same catalog. In particular, even unreliable observers would get aggressive weights if they were using a catalog with good statistics. Moreover, every time a new catalog is introduced, we would need to perform a statistical analysis of the stations using that catalog to determine the data weights, though it may take time before enough observations are available. For instance, all the astrometry reduced by using the Gaia-DR1 catalog \citep{Lindegren2016} was getting deweighted with respect to other catalogs just because Gaia-DR1 is a new catalog and does not have any statistics available. To deal with this problem, we dropped weighting rules only based on the star catalog, and only assigned rules to observatories we can fully characterize.

Also, the $\sqrt{N}$ scaling factor was introducing some issues. For instance, when only a pair of same-night observations, instead of the usual four, was submitted, the individual observations were overweighted. However, with only a pair there is less confidence that the astrometry is reliable. Moreover, the generic weight for CCD astrometry was at 1.5$''$ for the previous scheme and if four observations were submitted the weights were scaled at 3$''$, which appears to be too conservative. In the current scheme, we propose a constant weight up to four observations per night, and then deweight by a scaling factor $\sqrt{N/4}$. We selected a threshold of four observations per night since it is enough observations to make sure the object is real but not too many, which could cause problems due to unresolved systematic errors.

Tables~\ref{new_schema1} and \ref{new_schema2} give the weighting scheme based on the guidelines discussed above. For Catalina, Spacewatch and NEAT there is a time dependence that we account for (Table~\ref{new_schema1}). The remaining surveys did not exhibit statistically significant changes over time (Table~\ref{new_schema2}). Since we cannot fully characterize them, all the other CCD observations (less than 10\% of the whole CCD dataset) get a conservative weight at 1$''$ if the star catalog is known. If the catalog is unknown, there could be regional biases as large as 1$''$ \citep{Farnocchia15} and so observations are further deweighted at 1.5$''$. 

\begin{table*}[ht!]
\small
\begin{center}
\footnotesize
\caption{New astrometric weights for CCD observers with astrometric residuals dependent on epoch.}
\tabcolsep=0.11cm
\begin{tabular}{c|c|c}
\hline
MPC Code &$\sigma_{RA,DEC}$ & Epoch \\
\hline
\multirow{2}{*}{703} &1.0$''$ & $<2014-01-01$\\
 & 0.8$''$ & $>2014-01-01$\\
\hline
\multirow{2}{*}{691} &0.6$''$ & $<2003-01-01$\\
 & 0.5$''$ & $>2003-01-01$\\
 \hline
\multirow{2}{*}{644} &0.6$''$ & $<2003-09-01$\\
& 0.4$''$ & $>2003-09-01$\\
\hline
\end{tabular}
\label{new_schema1}
\end{center}
\end{table*}

\begin{table*}[ht!]
\small
\begin{center}
\footnotesize
\caption{New astrometric weights for the most active CCD asteroid observers.}
\tabcolsep=0.11cm
\begin{tabular}{c|c||c|c}
\hline
MPC Code &$\sigma_{RA,DEC}$ & MPC Code &$\sigma_{RA,DEC}$ \\
\hline
704&1.0$''$&C51&1.0$''$\\
G96&0.5$''$&E12&0.75$''$\\
F51&0.2$''$&608&0.6$''$\\
G45&0.6$''$&J75&1.0$''$\\
699&0.8$''$&other w/ catalog&1.0$''$\\
D29&0.75$''$&other w/o catalog&1.5$''$\\
\hline
\end{tabular}
\label{new_schema2}
\end{center}
\end{table*}

In addition to the 13 most prolific CCD surveys, there are some key NEO follow-up observers for which the 1$''$ weights would be far too conservative. Follow-up observers do not have as many observations as prolific survey and so it is harder to perform a statistical analysis that is as meaningful. Based on the experience and direct communication with some of them, we add the weighting rules described in Table~\ref{follow_up}. In some cases the usage of more accurate catalog (e.g., Gaia) warranted the usage of even tighter weights.

\begin{table*}[htb]
\footnotesize
\caption{New astrometric weights for selected NEO follow-up observers. Las Cumbres Observatories (LCO) are represented by following observatory codes: K92, K93, Q63, Q64, V37, W84, W85, W86, W87, K91, E10, F65.}
\begin{minipage}{\textwidth}
\begin{tabular}{c|c|c||c|c|c}
\hline
MPC   &\multirow{2}{*}{Catalog} & \multirow{2}{*}{$\sigma_{RA,DEC}$}&MPC  &\multirow{2}{*}{ Catalog }& \multirow{2}{*}{$\sigma_{RA,DEC}$} \\
Code & & & Code &  & \\
\hline
645&all&0.3$''$&Y28 & PPMXL, Gaia &0.3$''$\\
673&all&0.3$''$&568&USNO-B1.0, USNO-B2.0 &0.5$''$\\
689&all&0.5$''$&568 &Gaia &0.1$''$  \\
950&all&0.5$''$&568 &PPMXL &0.2$''$ \\
H01&all&0.3$''$&T09&Gaia& 0.1$''$\\  
J04&all&0.4$''$&T12&Gaia&0.1$''$\\
W84&all&0.5$''$&T14&Gaia&0.1$''$\\
G83\footnote{\label{aaa}Applies only to program code assigned to  M. Micheli, \url{ftp://cfa-ftp.harvard.edu/pub/MPCNewFormat/ProgramCodes.txt}}& UCAC-4, PPMXL &0.3$''$ & 309\footref{aaa}&UCAC-4, PPMXL  & 0.3$''$\\
G83\footref{aaa} & Gaia&0.2 $''$ &309\footref{aaa}&Gaia & 0.2$''$  \\
LCO&all&0.4$''$ &  \\

\hline
\end{tabular}
\end{minipage}
\label{follow_up}
\end{table*}

To validate the performance of the newly suggested weighting scheme, we performed a similar test to that described by \citet{Chesley10} and \citet{Farnocchia15}. Instead of MBAs, instead we used $\sim200$ NEOs with at least 5 apparitions.  We computed orbit solutions from subsets of 1 or 2 consecutive apparitions to be considered as predictions and to be compared to the long arc solution, used as truth. 

\begin{figure}[H]
\centering
        \includegraphics[width=0.48\textwidth]{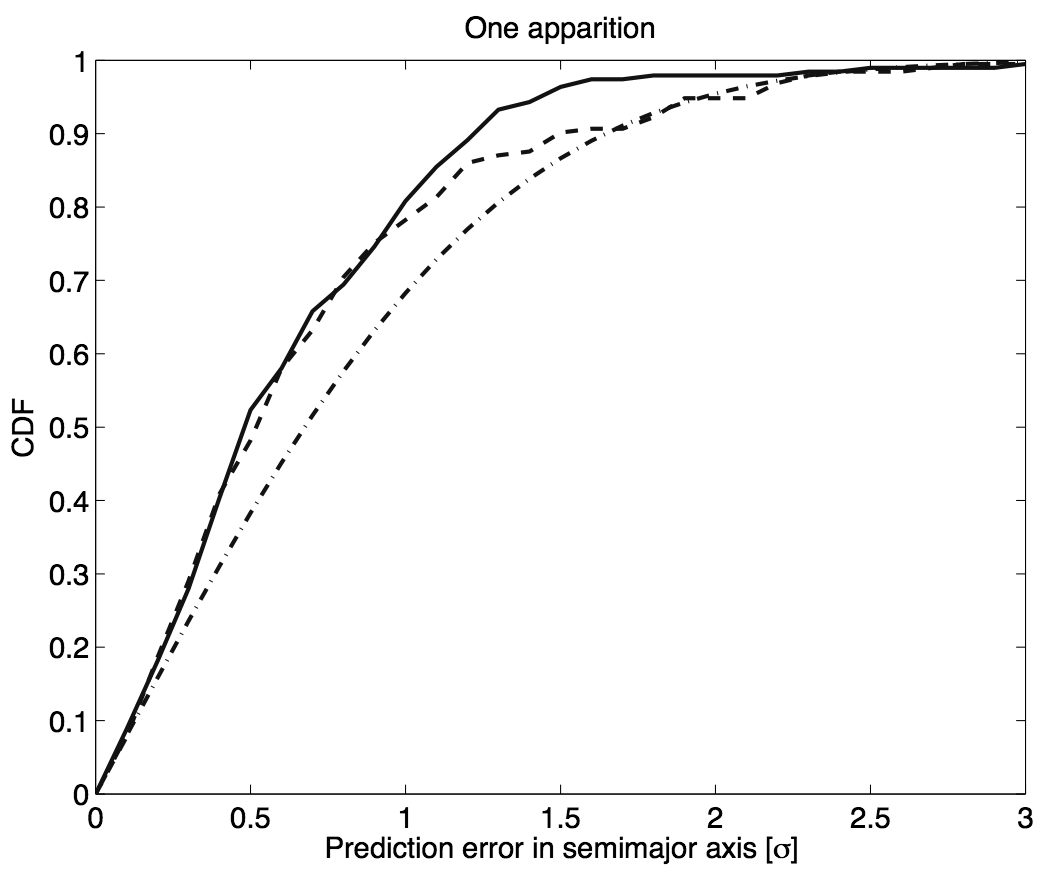}
         \includegraphics[width=0.48\textwidth]{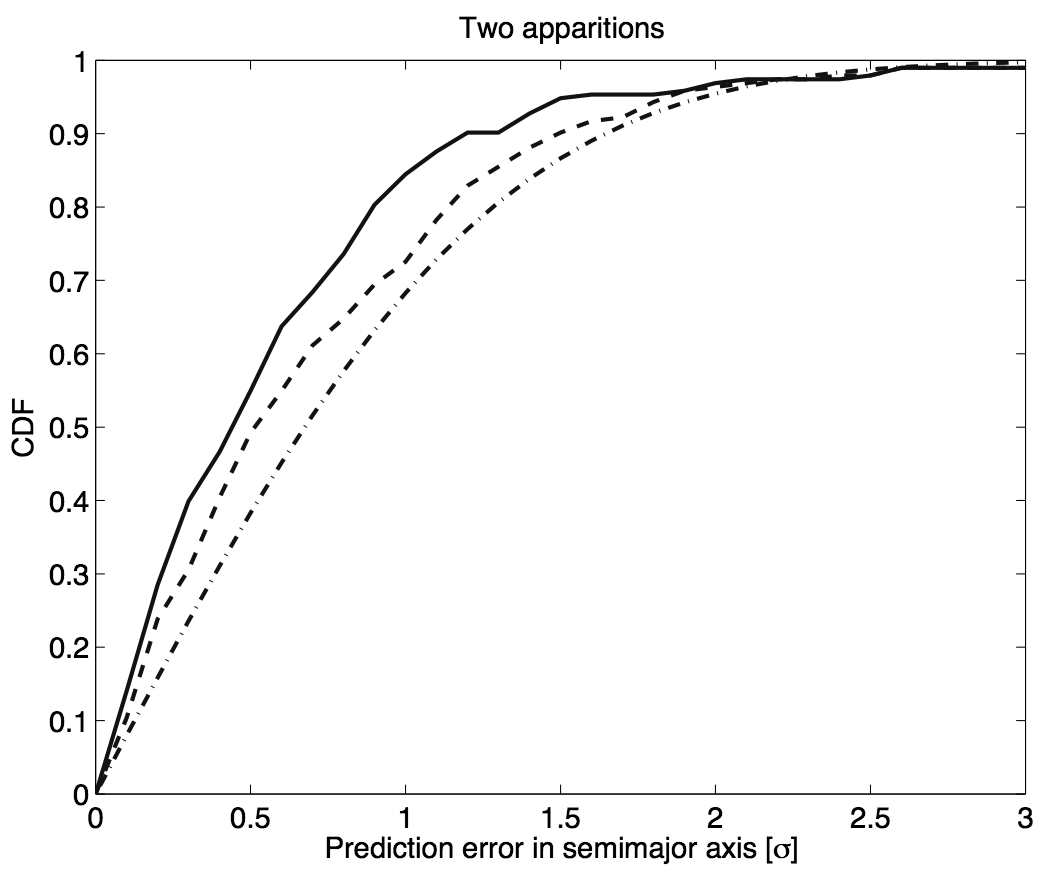}
         \includegraphics[width=0.48\textwidth]{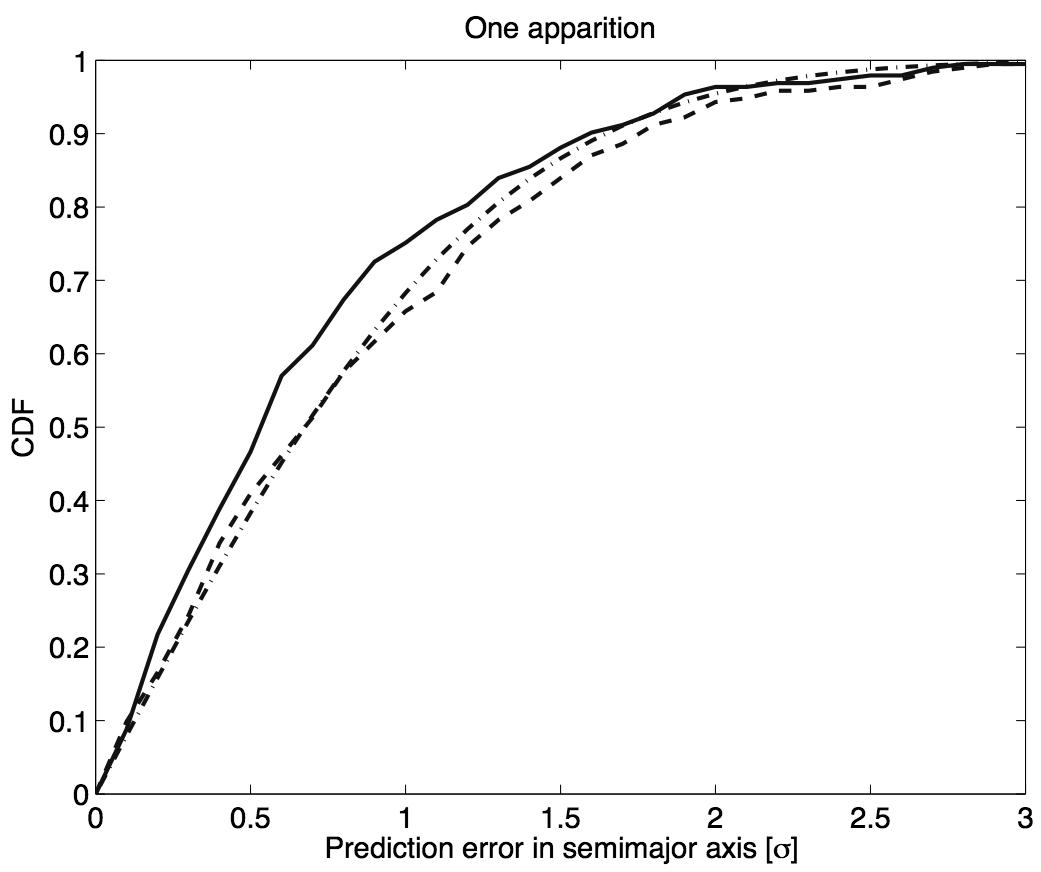}
         \includegraphics[width=0.48\textwidth]{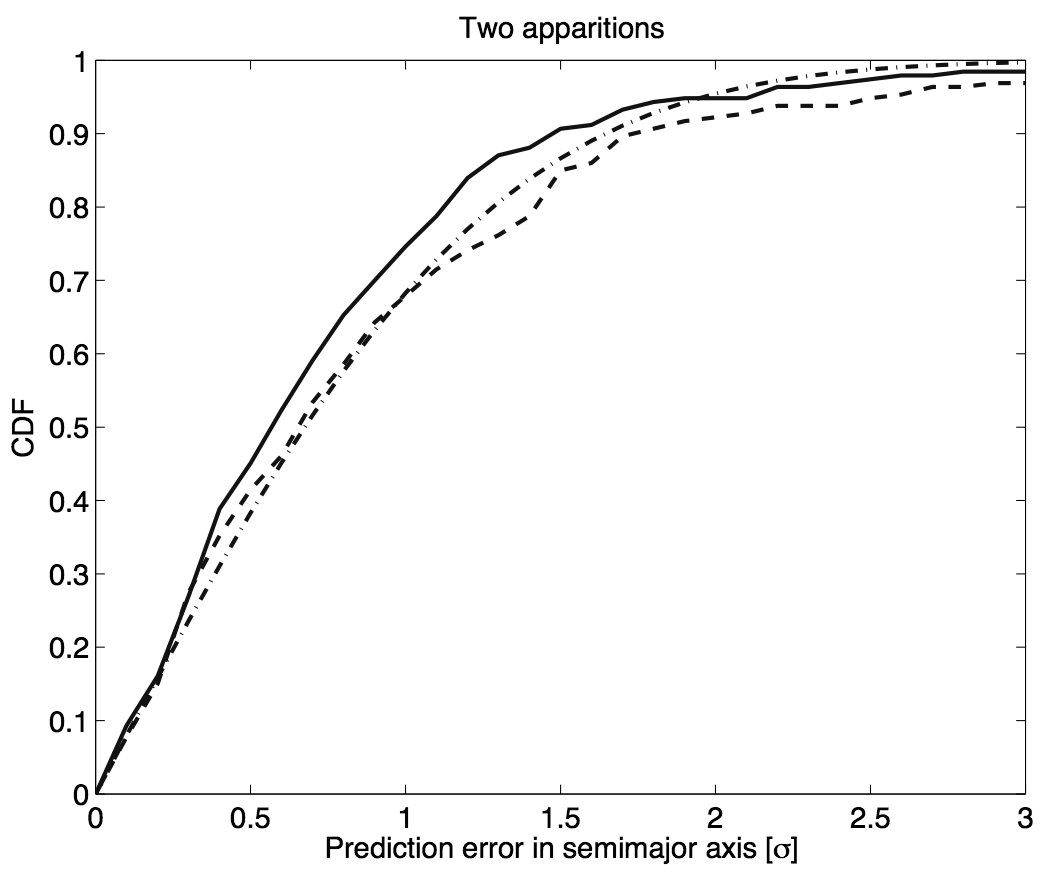}
\caption{Prediction error in sigma for semimajor axis for 200 NEO for the old (dashed) and new (solid) weighting scheme. Theoretical sigma CDF is depicted by a dash-dotted line. Upper row shows the most recent apparitions and the lower row shows the earliest apparitions (time span of 5 apparitions in total).}
\label{fig.sigma_a}
\end{figure}

To describe our results, we make the comparison in semimajor axis (Figure~\ref{fig.sigma_a}), as it is the key orbital parameter driving prediction uncertainties. We also did similar comparisons in 6-D orbital element space and Cartesian space, where the results and conclusions are the same. Generally, the cumulative density function (CDF) in semimajor axis is conservative when compared to a theoretical distribution by a factor of 1.5 or less. This conservative factor is likely due the fact we set the weights corresponding to the upper bound of RMS as a function of target's brightness. 

On the other hand, the new scheme seems to better capture the distribution tails than the old scheme and would therefore reduce the need of setting manual weights for outliers. We checked that the absolute prediction errors are essentially the same, which confirms the findings by \citep{Chesley10} and \citep{Farnocchia15} that weighting mostly affects prediction uncertainties, rather than the prediction itself.

\section{Data weights for non-CCD observations}

After CCD observations, the second most common type of observation in the current astrometry catalog is photographic. Even though this observation type dominates the pre-1995 astrometric dataset, we do not have enough data to perform a station-specific analysis as was done for CCD observations. 

\begin{figure}[H]
\centering
        \includegraphics[width=0.4\textwidth]{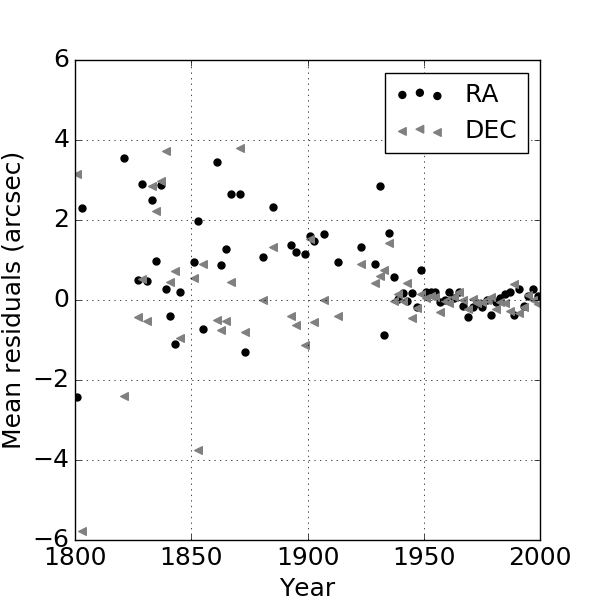}
         \includegraphics[width=0.4\textwidth]{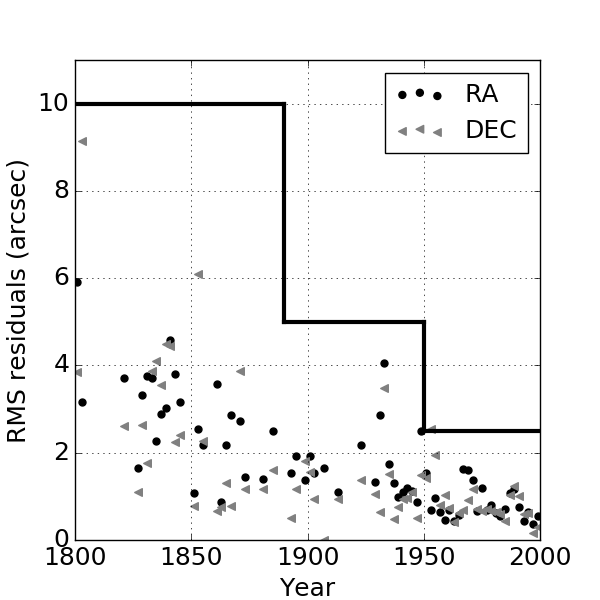}
                 \includegraphics[width=0.4\textwidth]{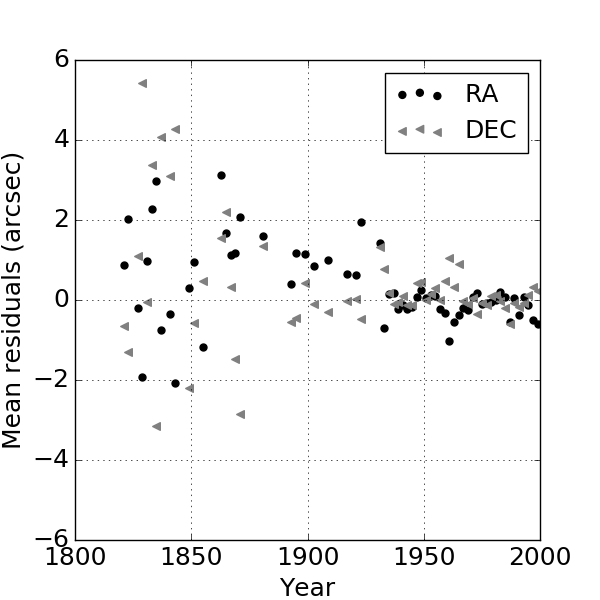}
         \includegraphics[width=0.4\textwidth]{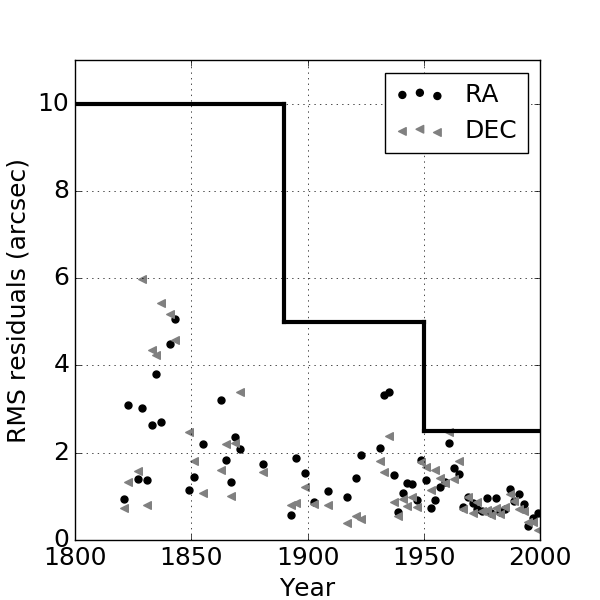}
\label{fig.ceres}
\caption{Mean (left) and RMS (right) residuals of photographic positions of Ceres (upper row) and Vesta (lower row) with respect to its orbit derived only from CCD astrometry. The step functions in the right panels correspond to the data weights reported in Table~\ref{new_schema3}.}
\end{figure}

To set the data weights for photographic astrometry, we selected Ceres and Vesta and computed an orbit using only their CCD, Hipparcos and transit circle observations. This orbit is accurate enough to predict their plane-of-sky positions at least 200 years in the past so we can compare it to photographic astrometry. Figure~\ref{fig.ceres} shows the RMS and mean RA and DEC pre-fit residuals for the photographic observations of these two objects, which we used to set up the weights (Table~\ref{new_schema3}). In any case, we generally advise caution when using old observations because of problems like UTC time not being defined, low precision in the reported astrometric positions, etc.

For other types of observations in the MPC dataset, there is even less information. For instance, brightness was often missing and in general the number of observations is small. Our data weights are derived from the RMS of the corresponding residuals and reported in Table~\ref{new_schema3}.

\begin{table*}[ht!]
\small
\begin{center}
\footnotesize
\caption{New astrometric weights for non-CCD observations. In the photographic category we include observations marked with ``P'' (or a blank), ``A" (converted from the B1950 frame) and with ``N" (normal place) by the MPC.}
\tabcolsep=0.11cm
\begin{tabular}{l|c}
\hline
Type of observation & $\sigma_{RA,DEC}$ \\
\hline
Photographic (before 1890-01-01) &10.0$''$ \\
Photographic (from 1890-01-01 to 1950-01-01) &5.0$''$ \\
Photographic (after 1950-01-01)&2.5$''$\\
 \hline
Occultations&0.2$''$\\
Hipparcos&0.2$''$\\
Transit circle&0.5$''$\\
Encoder&0.75$''$\\
Micrometer&2.0$''$\\
Satellite&1.5$''$\\
Mini-normal place&1.0$''$\\
\hline

\end{tabular}
\label{new_schema3}
\end{center}
\end{table*}

\section{Discussion}

The accuracy of asteroid orbits relies on that of the observations. In particular, the ephemeris uncertainties are a function of observation uncertainties. At present, no direct astrometric positional uncertainties are available for the MPC observation dataset. Nor is available information such as the detection SNR, which would allow one to infer what the astrometric uncertainties are. Therefore, our only option to quantify these uncertainties is to perform a statistical analysis of the astrometric errors.

We considered the 13 most productive CCD surveys, which account for more than 90\% of the overall astrometric dataset. For all of them we found that there is a significant dependence of the astrometric quality on a target's brightness. In particular, there is a clear quality degradation for faint objects, especially when near the survey's limiting magnitude, but also for the brightest objects likely due to saturation effects. This sensitivity is hard to capture by a simple metric such as the RMS of that survey's residuals.

We also found that the astrometric quality can change over time, which may correspond to telescope or camera upgrades, changes in the star catalog used for the astrometric reduction, etc.

Another line of investigation was the sensitivity of residuals on the rate of motion. Some surveys do not seem to properly handle detections of fast movers and the astrometric residuals significantly increase with the rate of motion. Difficulties in treating fast movers arise from the fact that detections are trailed. Also, for fast movers timing errors can result in significant position errors along the direction of motion. However, decoupling timing errors from unresolved systematic position errors is not trivial. Errors in clock timing can be mitigated by assuming timing uncertainties that would naturally map to plane-of-sky uncertainties.

Based on our statistical analysis we derived a new astrometric weighting scheme. The new scheme accounts for the dependence on target's brightness and observation epoch. For those CCD observations that we did not characterize, the weights are set in a conservative way. We found that the proposed weighting scheme is conservative by a factor of as much as 1.5. However, it better handles outliers and faint detections, and it reduces the need of manually setting the weights. Moreover, having some margin can be a good idea for applications such as impact monitoring or ephemeris support for space missions. The new scheme is being used by the JPL Solar System Dynamics orbit determination pipeline.

Future work will account for the Gaia star catalog \citep{Lindegren2016}. Its first release  has already proven useful by significantly reducing star catalog systematic effects \citep{Tholen2017}.
In 2018 the second release should become available and will include proper motions. At that point, it will possible to use the Gaia catalog to remeasure and improve past observations, and surveys may consider a massive reprocessing of their astrometry. 
Moreover, we plan to use the second release of Gaia to improve the \citet{Farnocchia15} debiasing tables and therefore subtract star catalog systematic errors to the current astrometric dataset.

Another upcoming significant change will be the new Astrometry Data Exchange Standard (ADES\footnote{\url{http://minorplanetcenter.net/iau/info/IAU2015_ADES.pdf}}). This format extends the one currently in use by the Minor Planet Center and allows observers to communicate valuable information such astrometric uncertainties and SNR that can be used to inform the weighting scheme.
However, even after the adoption of this new format, it is important to keep analyzing the performance of the various observers.
In fact, future weighting schemes will not necessarily convert the supplied uncertainties to data weights, at least until these uncertainties are proved to be accurate and to provide statistically consistent ephemeris prediction uncertainties.

Another aspect to be considered in the future is that of timing errors. This error component is important to deal with objects that have high rates of motion in the sky and for which timing errors can be significant or even dominate the astrometric position errors.

 \section*{Acknowledgement}

This  research  was  conducted  at  the  Jet  Propulsion  Laboratory, California Institute of Technology, under a contract with the National Aeronautics and Space Administration. \copyright\ 2017. All rights reserved.

\bibliographystyle{icarus.bst}\biboptions{authoryear}
\bibliography{references}

\end{document}